# Controlling the size distribution of nanoparticles through the use of physical boundaries during laser ablation in liquids

Kaushik Choudhury[1], R. K. Singh[2*], P. Kumar[1], Atul Srivastava[3#] and Ajai Kumar[2]

[1]IITB Monash Research Academy, IIT Bombay, Mumbai, India
[2]Institute for Plasma Research, Gandhinagar, India
[3]Indian Institute of Technology Bombay, Mumbai, India

E-mail: *rajesh@ipr.res.in, #atulsr@iitb.ac.in,

**Abstract**

A simple, yet effective method of controlling the size and size distributions of nanoparticles produced as a result of laser ablation of target material is presented. The method employs the presence of physical boundaries on either sides of the ablation site. In order to demonstrate the potential of the method, experiments have been conducted with copper and titanium as the target materials that are placed in two different liquid media (water and isopropyl alcohol). The ablation of the target material immersed in the liquid medium has been carried out using an Nd:YAG laser. Significant differences in the size and size distributions are observed in the cases of nanoparticles produced with and without confining boundaries. It is seen that for any given liquid medium and the target material, the mean size of the nanoparticles obtained with the boundary-fitted target surface is consistently higher than that achieved in the case of open (flat) targets. The observed trend has been attributed to the plausible role(s) of the confining boundaries in prolonging the thermalisation time of the plasma plume. In order to ascertain that the observed differences in sizes of the nanoparticles produced with and without the presence of the physical barriers are predominantly because of the prolonged thermalisation of the plasma plume and not due to the possible formation of oxide layer, select experiments with gold as the target material in water have also been performed. The experiments also show that, irrespective of the liquid medium, the increase in the mean size of the copper-based nanoparticles due to the presence of physical boundaries is relatively higher than that observed in the case of titanium target material under similar experimental conditions.

# 1. Introduction

In recent years, the method of laser ablation has found considerable attention among the researchers as one of the viable methods for the fabrication of nanoparticles. This method has been widely employed for fabricating metallic and polymeric nanoparticles [1–4]. The reason behind the upsurge in the use of this method is that it does not involve any harsh chemicals at any stage of the fabrication process. In addition, it is relatively simpler owing to the absence of any possible chemicals that are otherwise difficult to get rid of once the process of formation of nanoparticles is over. Laser ablation can be carried out in air/gas/vacuum or liquid ambient medium based on the desired end results. When laser ablation is carried out in air/gas/vacuum, it is mostly to let the ejected solid get deposited over a substrate and the method is known as pulsed laser deposition (PLD) [5,6]. On the other hand, when the process of ablation is carried out in liquid, the metal vapor cools down and forms nanoparticles [7,8].

It has been well established through a number of studies that the shape, size and morphology of nanoparticles play an important role in defining their importance in a large range of application areas that include paints, various heat exchange fluids to controlled drug delivery, catalysis and quantum computation [9,10]. Over the last few decades, the area of research has broadened from nanoparticles to nanostructures; encompassing not just spherical or nearly spherical nanoparticles but also a plethora of structures like prisms, cages, spindles, wires etc. [11–13]. The variation in the shape, size and/or surface morphology of the nanostructures leads to a change in their physical properties and response to a specific physical, chemical or biological stimulus [12,14]. This necessitates to develop a precise control over the shape and size of these nanostructures during their fabrication process itself. The parameters that are believed to influence the shape and size of the nanoparticles in the case of chemical synthesis include the concentration of surfactant, choice of precursor, rate of temperature variations etc. [11,15,16]. In addition, there are several process-specific parameters that lead to the formation of nanostructures with different shapes and sizes, based on the different mechanisms involved during the fabrication process.

In the context of laser ablation in liquids (LAL), one can chose media with different viscosities, densities, refractive indices etc.in order to control the size of the nanoparticles [17,18]. In some cases, choosing a reactive media also changes the size, size-distribution etc. of the resulting nanoparticles. These different methods have shown the capability of generating a wide range of sizes and size distributions of nanoparticles with different stoichiometry [19,20].

Researchers have also shown that parameters like laser wavelength, laser fluence and height of the liquid column also affect the size and size-distribution of the resultant nanoparticles [21,22]. In recent times, some research groups have come with very different and novel methods for controlling the size and the polydispersity of the nanoparticles that are fabricated by LAL. One such method is fabrication of nanoparticles by ablation of thin films fabricated over another substrate. It has been shown that the size of nanoparticles formed depends on the grain size of the material forming the film and also on the thickness of the film [23]. Working on the similar lines, the effect of laser fluence and the effect of cavitation bubble has been reported to be affecting the size and the polydispersity [24]. Also, it has been found that by making layers of thin films and by appropriately tailoring the ambient medium, it is possible to achieve core-shell kind of nanostructures of metal alloys. The thickness of the shell and the composition of the alloy could be altered by changing the thickness of the layers and changing the sequence, for a given liquid ambient [25].

The effect of geometrical confinement has also been found to affect the size and size-distribution of nanoparticles. Along these lines, some researchers have altered the thickness and width of the target material to achieve size control. By making the width of the target material smaller or larger than the beam size, different sized nanoparticles were obtained [26]. With this background, the present work explores the influence of physical boundaries placed in the vicinity of the ablation site on the size and the size distribution of nanoparticles. Experiments have been performed with flat target plate as well as the target plate fitted with confining walls, with the ablation site in the region equidistant from the physical boundaries (Al walls) and in between them. Copper (Cu) and Titanium (Ti) have been used as the target materials. The primary motivation of the present work has been drawn from one of our recent experimental studies wherein it has been reported that the presence of the physical boundaries in the close vicinity of the ablation site causes the reflection of the laser induced shockwaves and subsequently affects the medium density and the lifetime of the plasma plume [27]. In the present work, it has been demonstrated that the sizes of the nanoparticles produced due to the ablation of a metal target that has physical boundaries (walls) in the vicinity of the ablation site are significantly different from those produced due to the ablation of a metal target with no physical boundary. The experiments have been performed in two different liquid media with varying viscosities and densities. The nanoparticles produced from the two experimental configurations have been characterized using scanning electron microscopy (SEM) and the observations made have been discussed to develop an understanding of the possible reasons that lead to the variations in the size of the nanoparticles produced following these two different approaches. The primary findings of the present experimental study are expected to pave a way for finding a relatively simpler, yet an effective method that can potentially be employed to achieve control over the size of the nanoparticles fabricated by the means of laser ablation.

## 2. Experimental setup

The complete experimental setup employed in the present work has been shown schematically in Figure 1. The fundamental (1064 nm) of Nd:YAG laser (*Quantel Q-smart* 850), with a pulse duration of 6 ns has been used as the ablation beam. The ablation beam has been focused using a plano-convex lens (*f* = 175 mm) and is made to hit the target surface at normal incidence. The energy of the laser beam has been set to 87 mJ and the beam has been focused to achieve a spot-size of 0.5 mm. For all the experiments reported in the present work, the spot-size and the laser energy have been maintained constant. A small fraction of the beam was split and let to fall on an optical energy meter to ensure that the laser fluence is constant during the course of the experiments.

Copper and titanium plates (99.9% purity) with dimensions of 25 mm × 25 mm × 2 mm have been used as the target plates. Before the start of the experiments, the surfaces of the target plates were polished and thoroughly rinsed in acetone to avoid impurities due to oxide formation. Two different target geometries have been used for both the target materials. The schematics of these two configurations of target plates have been shown in Figure 2. Figure 2(a) shows the flat plate target without any physical boundary and Figure 2(b) shows the target plate fitted with two Aluminum confining walls that are 10 mm apart from each other. The ablation was carried out in such a way that the point of ablation remains equidistant from both the confining walls. The Al walls were put on the target surface to reflect back the laser-induced shockwaves towards the ablation site (plasma plume). Phenomena such as the reflection of laser produced plasma-induced shockwaves from the confining boundaries and their characteristics in liquid media of different densities have been studied and reported in detail in our recent articles [27].

The LAL experiments have been performed in a custom made glass cuvette of dimensions 60 mm × 60 mm × 60 mm. The top of the container was left open. The chosen dimensions of the cuvette made sure that the total travel time that the shock waves (generated due to the ablation of the metal) take in travelling from the point of ablation to the walls of the container and back again to the ablation point (after getting reflected from the container walls) is significantly greater than the thermalisation time of the plasma produced during the process. This condition needs to be satisfied to avoid any plausible effect(s) of the reflected shockwave on the plasma plume, with the flat plate target geometry, i.e. in the case where there are no ablation boundaries. If this condition is not met, *viz.* in the case of a smaller-sized container, then the reflected shockwave will affect the process of thermalisation of the plasma. Hence, the process of formation of nanoparticles will get influenced and one will not be able to achieve what may be termed as LAL of a flat target without any confinement. The ablation chamber has been kept on a translation stage (20 μm precision) and the target was shifted constantly to avoid any unwanted effects arising due to the formation of crater or surface modification on the target surface.

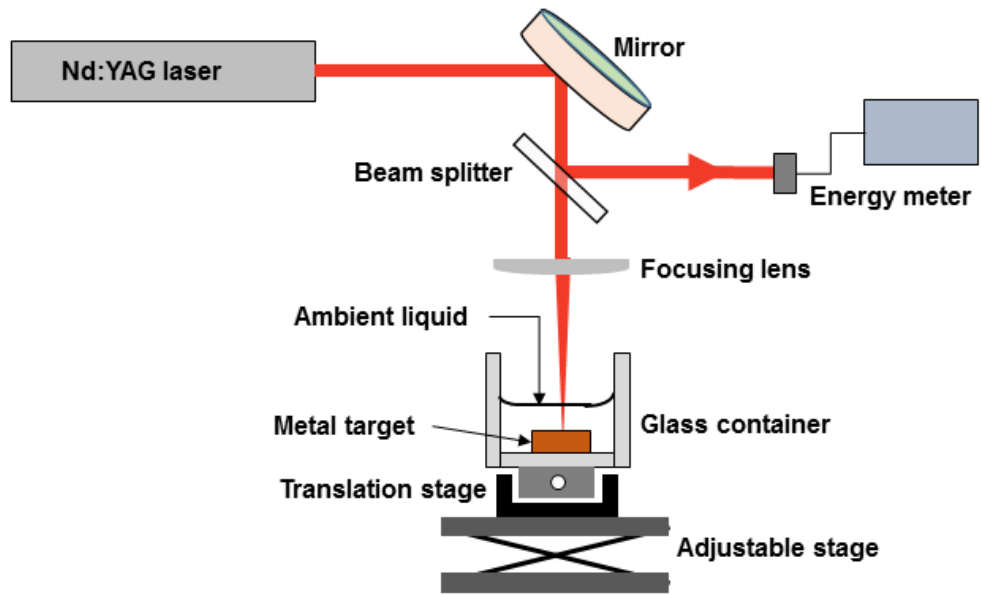

**Figure 1: Schematic diagram of the experimental setup**

To understand the effect of the confining boundaries, four sets of experiments have been performed. In this direction, two different target materials and two different ambient liquids were chosen. Water and isopropyl alcohol (IPA) have been chosen as the ambient liquids. The reason for selection of IPA and water as the two liquid media has been based on the fact that their viscosities and densities are significantly different from each other and hence significant contrast in the effect of confinement may be expected. At the same time, the refractive indices of the two liquids differ slightly and hence the transmission and reflection from the liquid target interface do not differ significantly for a chosen target material. It enables us to assume that the laser fluence is constant for a specific target material in different ambient fluids. Two sets of experiments have been performed with copper and titanium targets (with and without confining walls) in water ambient and two other sets of experiments have been performed with the same targets but with IPA as the ambient medium. In each case, the height of the liquid column was maintained at 10 mm with reference to the top surface of the target surface. Ablation of the target surface has been achieved by firing the laser beam at 5Hz for 30 minutes (9000 shots). The laser fluence was maintained at 11 J/cm$^2$. This time interval has been chosen so as to make sure that there are not too many nanoparticles in the suspension so as to avoid any possibility of agglomeration. On the other hand, the number is high enough that is required for characterizing the produced particles for size distribution using SEM. The extent of the physical boundaries were just enough to confine the ablation spot and enable enough room for the nanoparticles to drift with time.

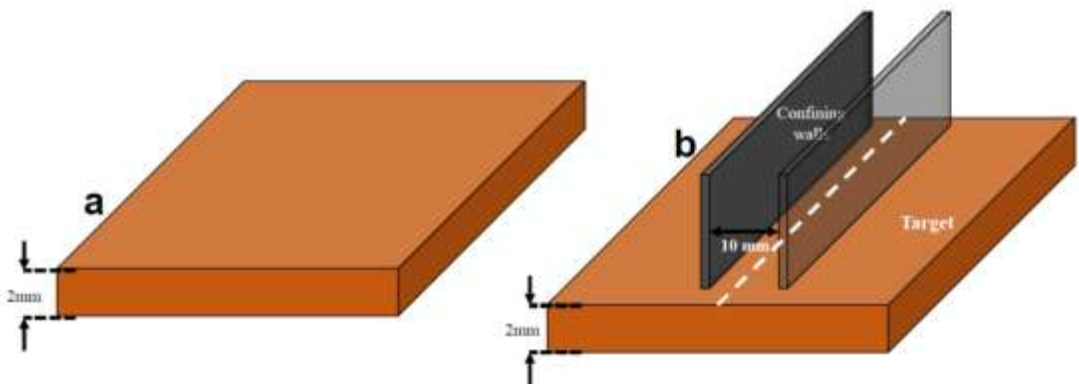

**Figure 2: Schematic representation of the two configurations of the target plate. (a) Flat (open) target and (b) target fitted with confining physical boundaries (Al-wall) on either sides of the ablation site.**

## 3. Materials and methods

Laser ablation in liquid (LAL) has been carried out with two different target materials immersed in different liquid ambient media under constant laser fluence and height of the liquid column. The as prepared suspensions carrying ablated nanoparticles have been scanned using SEM (*Zeiss-Merlin*). The instrument has a resolution of 0.7 nm at 15 kV and 1.2 nm at 1 kV. For the preparation of samples for SEM characterisation, a few drops of the suspension was dried over a cleaned Al foil, which was later put on the sample holder using a carbon tape. Before preparing the samples, the suspension was sonicated well to avoid agglomeration and ensure homogeneity. The SEM images were later analysed using image processing codes to retrieve the particle size distribution.

The images having nanoparticles were processed using image processing toolbox in MATLAB. The images recorded using scanning electron microscopy were processed to enhance the quality of images to enable particle detection through Hough Transform. The images were pre-processed based on the difference in the contrasts between the foreground and background. Here, the foreground is formed by the nanoparticles (single or aggregated) and background is the substrate onto which nanoparticles were deposited (aluminium foil). Due to the differences in the height among the nanoparticles or nanoparticle clusters with respect to the substrate, there are differences in the intensity values between foreground and the background. Although, the intensity among the nanoparticles has also got some variations, depending upon the size distribution, yet the intensity value for nanoparticles is always greater than the intensity value of the background substrate. These differences in intensity values of substrate and nanoparticles were amplified to enhance the contrast of the image. This was achieved by adjusting the parameters manually by looking at the intensity distribution in the images. Thereafter, the images were filtered through top-hat filter. It increased the intensity of brighter pixel with respect to dark background through morphological openings of the image using a structuring element. The shape and size of the structuring elements were manually tuned for each image to enable enhancement in contrast of image while preserving the features of interest (nanoparticles) in the images. After, enhancement in the quality of the images, Hough Transform (HT) was applied on the processed images. The parameters of the command were fine tuned for each image to prevent an overestimation or underestimation of diameter and centre of the circle (2D projection of spherical nanoparticles). The HT converted 2D circular objects on an image to an accumulator matrix in a Hough space represented by three parameters; diameter, co-ordinates of centre of a circle. Thereafter, a 'voting' is performed to determine a local maxima in an accumulator matrix. Following this process, the maxima is mapped to the image space to locate the circle in an image. The diameters of the circle obtained in pixel were converted to real size values through appropriate scaling with scale bar provided on the SEM images. The images were manually inspected to verify the accuracy of the algorithm and its parameter values in detection of the nanoparticles sizes. Thereafter, the data was processed to generate histograms representing the size distribution of the nanoparticles. The above process was followed for each image and algorithm parameters were tuned to determine the size distribution of nanoparticles.

Figure 3 shows a set of representative images to depict the process of extraction of sizes of the nanoparticles. Figures 3 (a) and (b) show the raw and the contrast enhanced images, respectively. It can be clearly seen that Figure 3(b) has a better contrast than Figure 3(a) and this contrast difference is between the features of interest (nanoparticles) and the background (substrate). On this enhanced image HT was applied and the circular features were selected. Figure 3(c) is the image that shows the outline of the nanoparticles as generated after applying Hough transform. Figure 3 (d) is the magnified version of Figure 3 (c) to show the effectiveness of the algorithm qualitatively. Once this process is over the sizes of these circular features were calculated and tabulated by converting pixel dimensions into the physical dimension using the scale bar of the SEM images. The parameters of the filters (used for contrast enhancement) and HT were so chosen that the features of interest remain intact. The same process was applied on multiple SEM images taken for the same sample at different position of the substrate. The size distribution obtained from the multiple images was tabulated and the mean size of the nanoparticles were obtained. A flow-chart of the algorithm has been shown in Figure A-1 in the appendix.

The method of extraction of particle size and size distribution is limited by the accuracy of the algorithm. The algorithm is based on Hough transform and can only detect simple geometric structures, i.e. lines, circles etc. As the complexity increases the computational efficiency and accuracy is compromised. In the present set of data circles have been detected and the minimum detection limit is 3 pixels that satisfy the parametric equation of the circle.

A detailed discussion on the findings has been made in the next section under two headings, *viz.* the effect of physical boundaries in water ambient and the effect of boundaries in IPA ambient. Since the focus of the work is on the investigation of the presence of physical barrier on the size distribution of nanoparticles, the ambient media have been kept similar while analysing the data and hence this classification. The viscosities of water and IPA at 25$^o$C are 0.89 cP and 2.04 cP, respectively, significantly different from one-another [28]. The densities of water and IPA at 25$^o$C are 0.997 g/cm$^3$ and 0.786 g/cm$^3$ respectively. Copper and titanium have been chosen as the target materials because they respond differently to the applied laser pulse [29]. The compositions of Cu and Ti nanoparticles formed in water and IPA ambient have been thoroughly reported and investigated by many researchers [20,30,31]. The refractive indices (RI) of Cu and Ti are significantly different at the ablation wavelength (1064 nm) and hence for any given liquid medium, these materials exhibit appreciable differences in their reflectivity and transmittivity. The values of refractive indices have been tabulated below in Table 1 [28]. The corresponding reflectivity values derived from these RI values have been tabulated in Table 2.

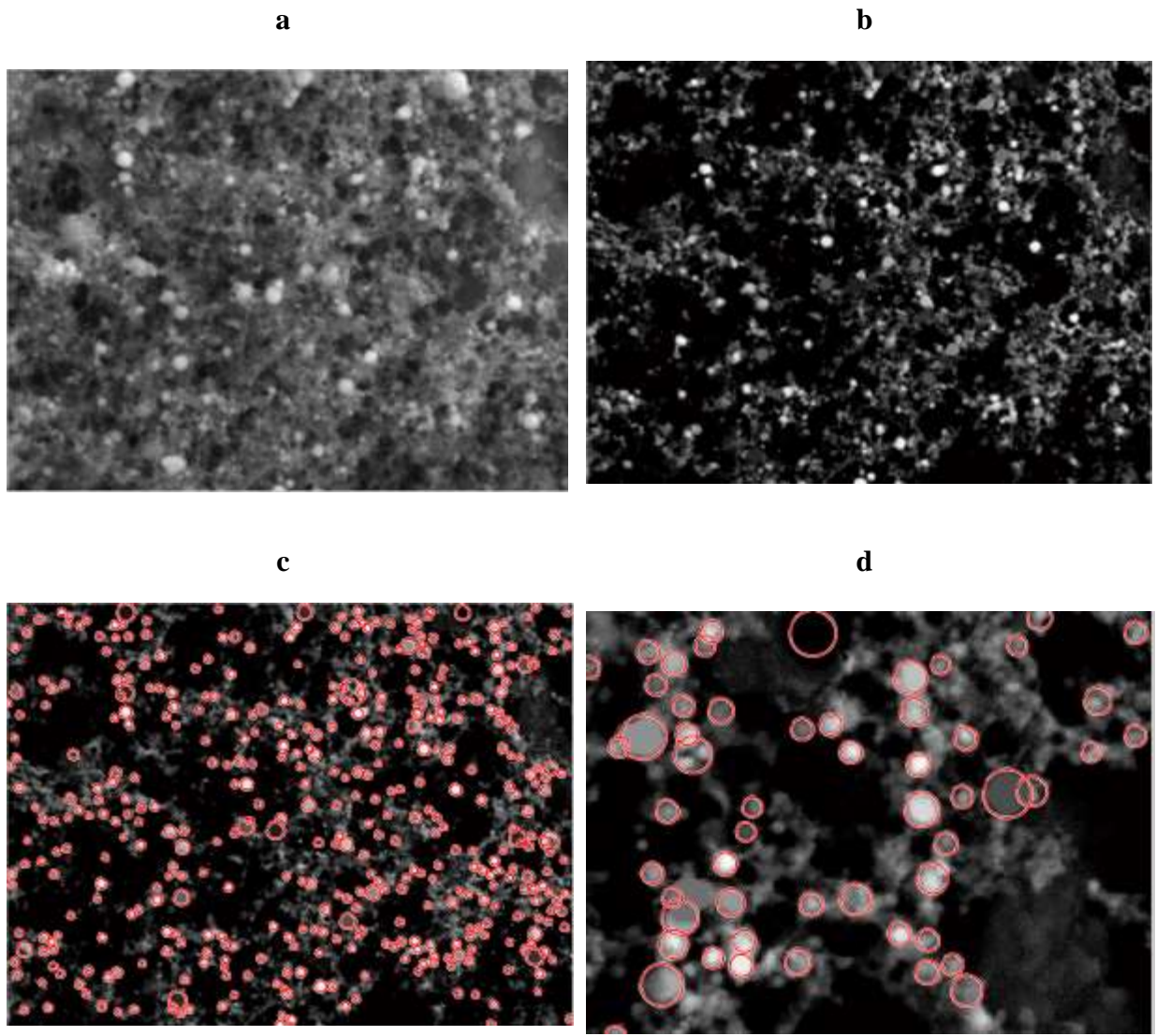

**Figure 3: Representative images to show the different stages of the process of extraction of the size of nanoparticles from the SEM images**

| Table 1: Refractive indices of the target materials and ambient liquids | | | | |
|---|---|---|---|---|
| Material | Copper | Titanium | Water | IPA |
| RI | 0.37861 | 3.4654 | 1.332 | 1.3514 |

| Table 2: Reflectivity values at the interfaces | | |
|---|---|---|
|  | Copper | Titanium |
| Water | 0.310627 | 0.197757 |
| IPA | 0.316185 | 0.192616 |

## 4. Results and discussion

Primary findings of the laser ablation experiments conducted with two target materials immersed in liquid media have been discussed in the present section. Results have been categorized primarily into two different sections; the first section discusses the effects of the presence of confining boundaries when the target is kept in water ambient, while the observations made with IPA as the liquid environment have been presented in the second subsection. For each liquid medium, the experiments have been conducted with two types of target materials i.e. copper and titanium.

4.1 Experiments in water ambient

As discussed earlier, the size of nanoparticles produced by the method of laser ablation depends on a range of parameters that include laser fluence, ambient medium, physical properties of the target material etc. Effect of many of these parameters on the size distribution of the nanoparticles have been investigated by various researchers in the past, both, experimentally as well as theoretically [13,29,32–34]. In the present discussion, all such parameters (e.g. laser fluence, ambient medium etc.) have been maintained constant for each set of experiments conducted such that any change occurring due to the presence of physical barriers may be clearly seen. Thus, the changes that are observed in the sizes and size distribution of nanoparticles produced due to laser ablation can be attributed to the role played by the presence of physical barriers on the target surface.

Under these conditions, the laser ablation has been carried out on the target surface that is either flat (no physical barrier) or is confined by the physical boundaries (walls) on both the sides of the point of laser ablation. Following the methodology discussed in the previous section, samples in the form of suspensions of nanoparticles have been subjected to SEM analysis and the corresponding SEM images have been shown in Figures 4 and 5. SEM images for copper targets, both flat as well as that fitted with confining walls, and the corresponding plots for size distribution of the particles have been shown Figure 4. The same studies have been carried out with titanium as the target material and the results have been shown in Figure 5. In both the figures, the subfigures (a) show the SEM images of the nanoparticles produced due to the ablation of the flat target plate (without barriers), while the subfigures (c) correspond to the case of copper (Figure 4(c)) and titanium (Figure 5(c)) target plates with confining walls. The subfigures (b) and (d) are the particle size distribution plots. The SEM images shown have been quantified and the results have been presented in the form of the size distribution plots in Figure 4(b) and 4(d) (for copper target in water) and Figure 5(b) and 5(d) (for titanium target in water). These size-distribution plots clearly highlight the possible influence of the confining boundaries in the form that, for any given target material type, the average size of the nanoparticles produced due to the ablation of the flat plate is relatively smaller than that achieved in the case of target surface fitted with physical boundaries. These observations have been quantified in terms of the mean size of the nanoparticles produced with water environment and summarized in Table 3. It is to be seen from the table that, for any target material, the size of the most abundant nanoparticles is consistently higher in the case of the ablation target fitted with the physical boundaries as compared to the case of the flat target. For instance, a percentage increase of nearly 66% in the mean size of the copper-based nanoparticles is to be seen as the flat (open) target is replaced by the one that is fitted with the physical boundaries. Similar trends are to be observed in the case of titanium substrate, though the percentage increase in the mean size of the nanoparticles is not as significant as was observed with copper as the target material and here the increase in the nanoparticle size is restricted to $\approx 17.5\%$.

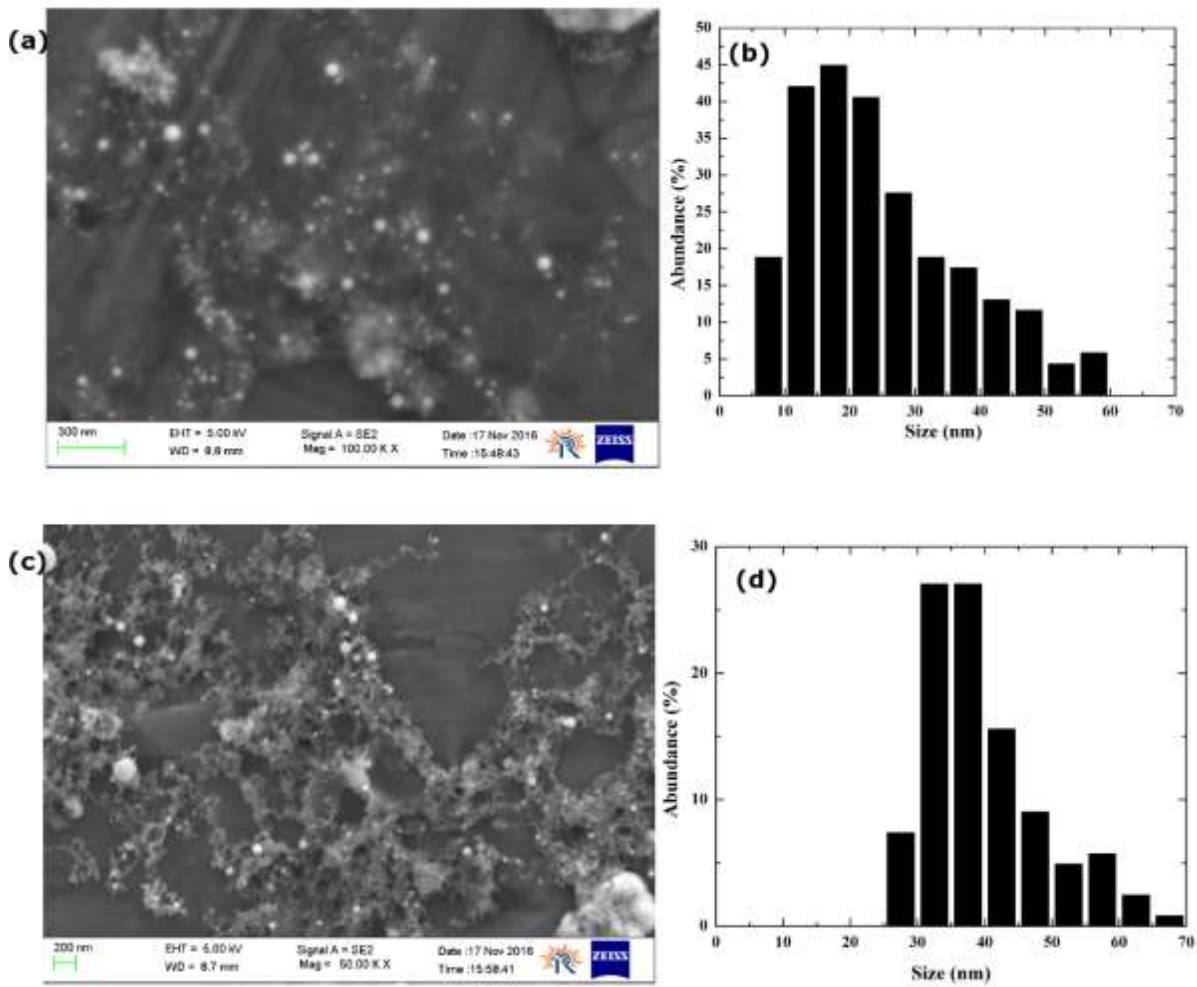

**Figure 4: SEM micrographs of the nanoparticles produced as a result of ablation of a flat copper target (a) and copper target fitted with confining physical boundaries (c). Corresponding size-distributions of the resultant nanoparticles have been presented in (b) and (d). (Ambient liquid medium: Water)**

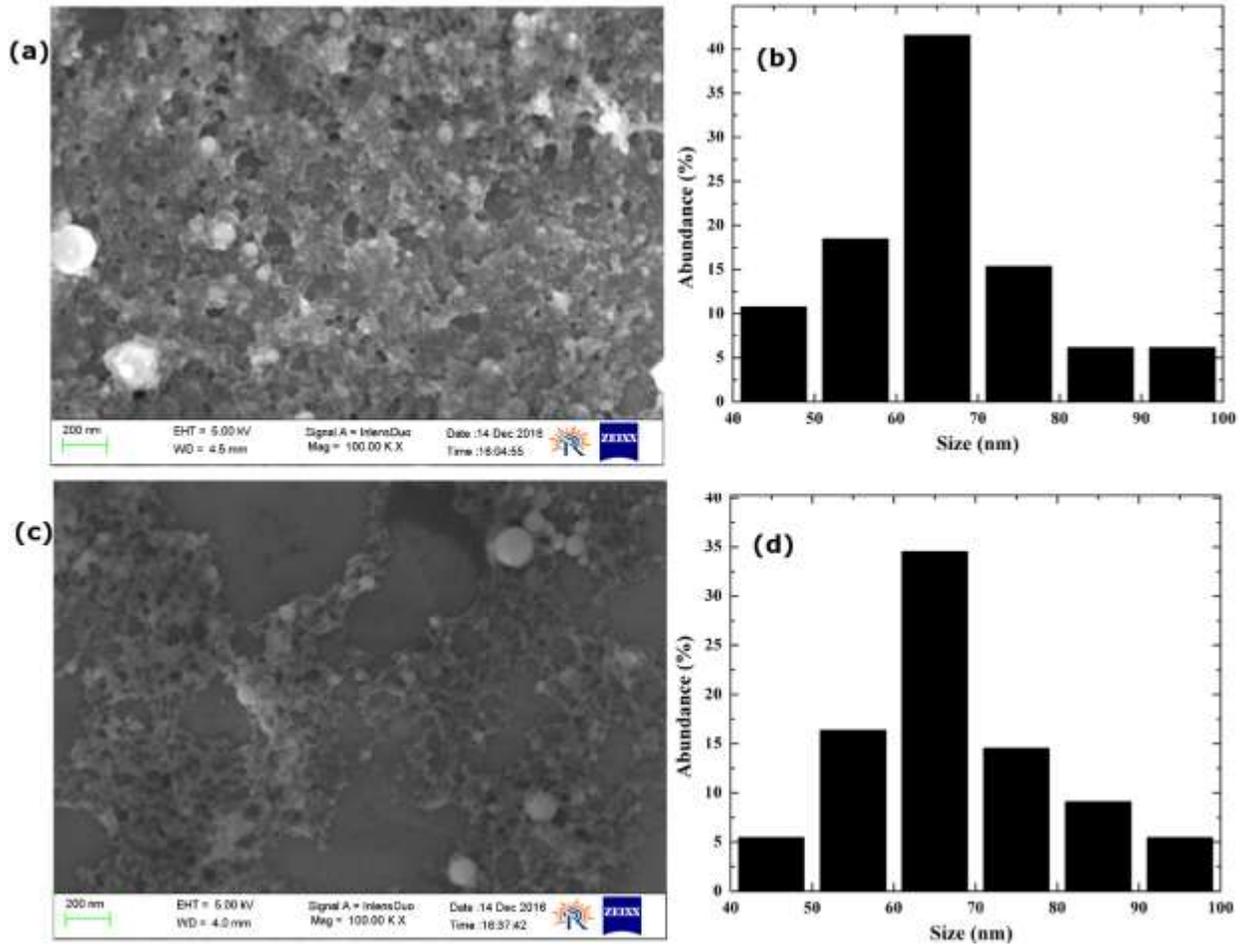

**Figure 5:** SEM micrographs of the nanoparticles produced as a result of ablation of a flat titanium target (a) and titanium target fitted with confining physical boundaries (c). Corresponding size-distributions of the resultant nanoparticles have been presented in (b) and (d). (Ambient liquid medium: water)

The above observations made in the form of an increase in the mean size of the nanoparticles produced with target fitted with physical boundaries may be explained on the basis of the rate at which the thermalisation of the plasma plume takes place. It has been well documented in the available literature that the process of formation of metallic nanoparticles by laser ablation takes place due to the thermalisation of the plume. Thus, the rate at which the plasma plume cools down plays an important role in this entire process [35]. A very high cooling rate of the plasma plume is expected to result into smaller size of the nanoparticles produced. This is owing to the fact that the formation mechanism of nanoparticles after the completion of the ablation process depends on two co-occurring processes, namely, nucleation and particle growth. The nucleation process starts as soon as the ablated species come together, followed by the coalescence of these nuclei that in turn leads to the formation of bigger particles. If the thermalisation is fast enough, the nucleation process dominates, leading to the formation of smaller nanoparticles. On the other hand, if the plume lifetime is longer, i.e. if the process of thermalisation takes place over a longer period of time, the nuclei produced as a result of material ablation get enough time to come closer and form bigger particles. In the context of the present work, the presence of the physical boundaries (Al walls), on either sides of the ablation site on the target surface, leads to the reflection of the shockwaves towards the site of laser ablation, thereby compressing the ambient medium. The subsequent increase in the pressure leads to a longer plasma thermalisation time, which causes the enhancement in the sizes of the nanoparticles [27,35].

Another interesting observation that is to be made from Figures 4 and 5 is in the form of the larger mean size of the titanium nanoparticles in comparison to that achieved with copper as the target material for any given configuration. The observed differences are to be attributed to the relative differences in the refractive indices of these two materials. Thus, for the given ablation wavelength (1064 nm) employed in the present set of experiments, the absorption coefficients of the two target materials are expected to be different. Due to the differences in the absorption coefficients, the fraction of the total energy of the incident laser pulse that causes the ablation of the target material will be different and hence the extent of ablation. This is because, the amount of mass ejected and the temperature of the plasma will be different for the two materials[1].

The experiments performed in water ambient with copper and titanium show similar trends, but there is a probability that these metals will form oxides. Hence the sizes or size-distribution observed may also have a contribution from oxide layers. In order to ensure that oxide formation is not a predominant factor, the experiments were also performed with gold (99.9% purity) as target material. The possibility of formation of oxide in the case of gold is quite minimal and hence the sizes observed may be considered as the size of the metallic nanoparticles. Therefore, the difference in the sizes of the nanoparticles that result from the ablation of a flat target plate and those formed due to the ablation of metallic plates fitted with physical barriers can be attributed to the presence of the physical barriers. The results obtained with gold targets ablated under water have been shown in Figure 6. Figures 6(a) and 6(c) show the SEM images of the gold nanoparticles formed by ablating a flat target and target fitted with walls, respectively. Figures 6(b) and (d) show the corresponding size distributions.

---

[1] It is pertinent to note here that factors such as electron-phonon coupling strength, surface energy, melting and boiling point of the material etc. may also influence the size and the size distribution and may contribute towards the observed differences in the produced nanoparticles [29]. However, the primary findings of the present experiments have consistently revealed the plausible role(s) of the confining physical boundaries in increasing the mean size of the nanoparticles produced as a result of laser ablation of the target material (copper and titanium).

The SEM images and the corresponding size distributions shown in Figure 6 clearly indicate that the size of the gold particles formed due to the ablation of metal target fitted with walls is significantly bigger than their counterparts that are produced by ablating a flat target. It is, therefore, concluded that the presence of barrier leads to prolonged thermalisation of the plasma plume and hence formation of bigger nanoparticles. The results have been tabulated in Table 3.

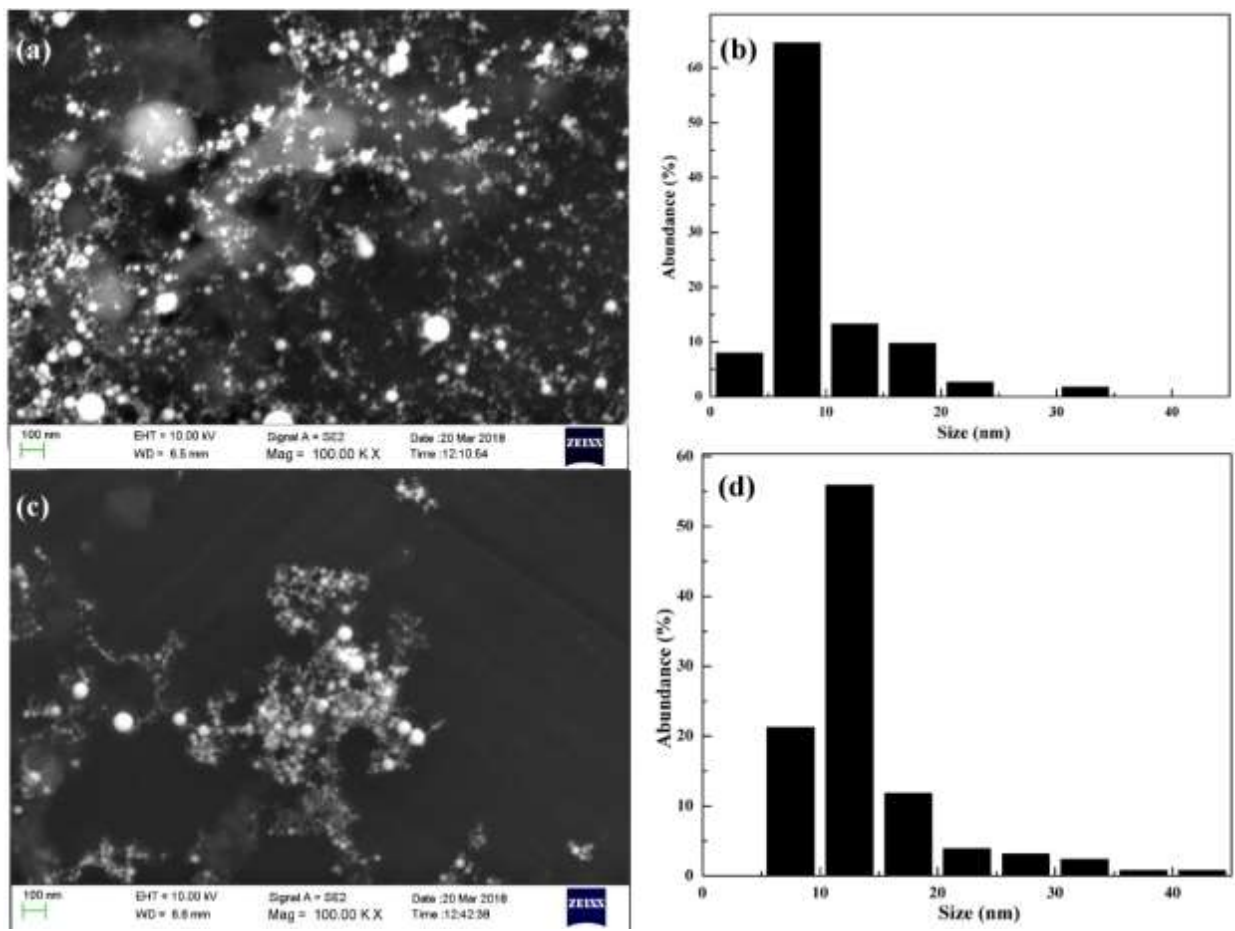

**Figure 6: SEM micrographs of the nanoparticles produced as a result of ablation of a flat gold target (a) and gold target fitted with confining physical boundaries (c). Corresponding size-distributions of the resultant nanoparticles have been presented in (b) and (d). (Ambient liquid medium: water)**

**Table 3: Comparison of the mean size of the nanoparticles produced with flat target to those obtained with the target fitted with physical boundaries (Ambient liquid medium: Water)**

| Material | Mean size of nanoparticles | | % Increment |
|---|---|---|---|
| | Flat Target | Target with boundary | |
| Copper | 25.5±13 nm | 42.4±13 nm | 66 |
| Titanium | 67.6±16 nm | 79.4±32 nm | 17.5 |
| Gold | 9.15±5 nm | 13.7±6 nm | 49.7 |

4.2 Experiments in IPA ambient

Observations made on the possible effects of the confining boundaries on the resultant sizes of the nanoparticles produced in isopropyl alcohol (IPA)-based liquid medium have now been presented and discussed. Figures 7 and 8 respectively show the SEM images and the respective particle size distributions due to the laser ablation of copper and titanium target surfaces immersed in IPA ambient. Subfigures (a) and (c) in both the figures respectively show the SEM images of the nanoparticles formed with flat target (a) and target fitted with physical boundaries (c). The respective particle size distributions of the produced nanoparticles have been shown in subfigures (b) and (d). Irrespective of the material of the target plate, smaller-sized nanoparticles are to be observed in the case of flat targets while the nanoparticles are relatively bigger in size when produced with confining walls placed on the target surfaces. These results follow the similar trend as was observed in the case of water as the liquid ambient medium presented in the previous section. However, a closer observation of the mean size of the nanoparticles summarised in Table 4 reveals that the copper nanoparticles produced under IPA liquid ambient are relatively larger in size as compared to their counterparts obtained with water as the ambient liquid medium (Row 1 of Table 3). On the other hand, the IPA ambient-based titanium nanoparticles are smaller in size in comparison with the size of the titanium nanoparticles produced with water as the liquid medium (Row 2 of Table 3). These differences may be attributed to the differences in the thermos-physical properties of the two target materials considered in the present set of experiments [29].

Also, a comparison of the mean particle sizes for the two materials presented in Tables 3 and 4 shows that in any given liquid ambient, the difference in the size of the copper nanoparticles produced from flat target and walled target is much higher as compared to that observed in the case of titanium nanoparticles produced under similar conditions. Such a trend may be attributed to the differences in the initial plasma temperatures of the two target materials and the amount of mass ejected per shot for a given set of operating parameters maintained during the course of the experimental runtime.

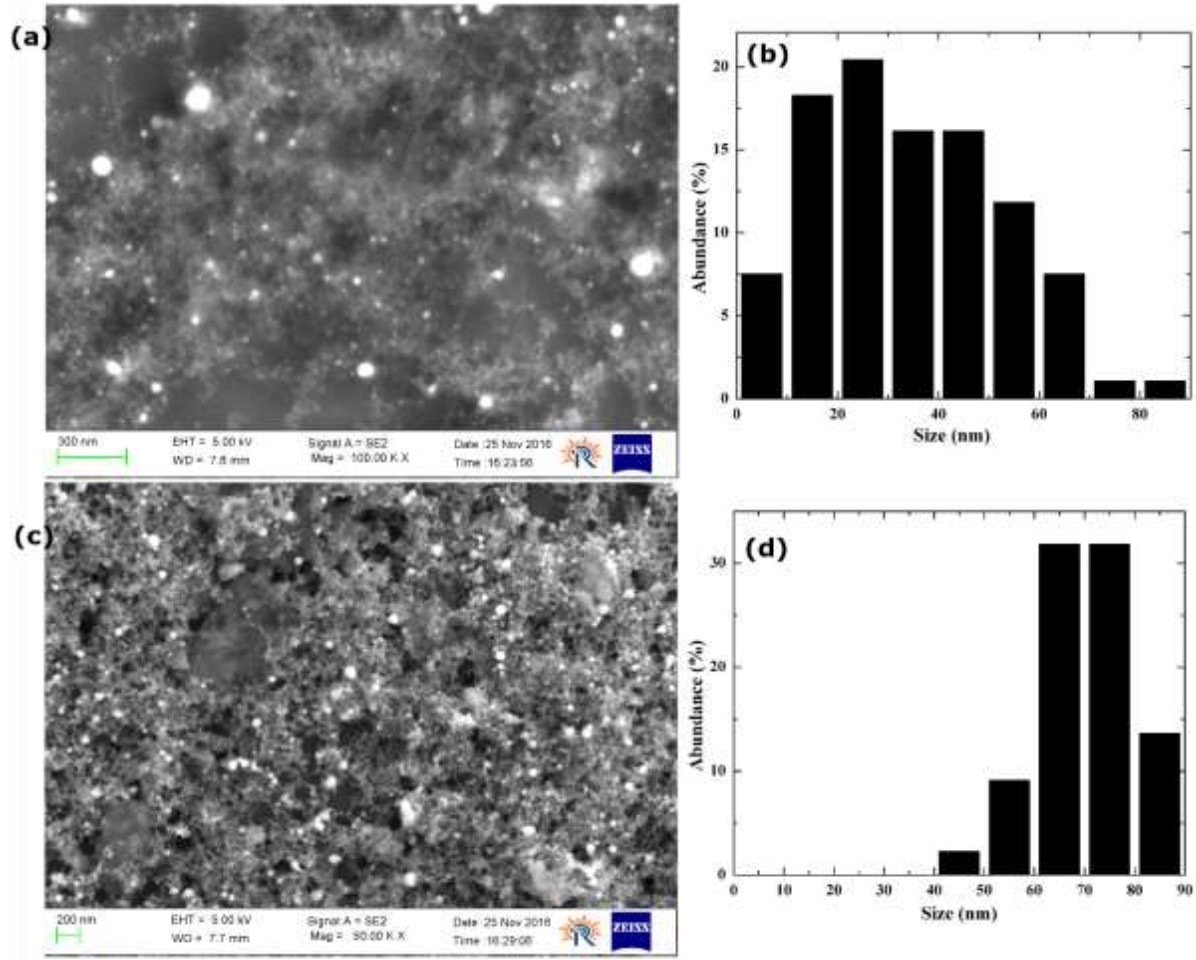

**Figure 7: SEM micrographs of the nanoparticles produced as a result of ablation of a flat copper target (a) and copper target fitted with confining physical boundaries (c). Corresponding size-distributions of the resultant nanoparticles have been presented in (b) and (d). (Ambient liquid medium: IPA)**

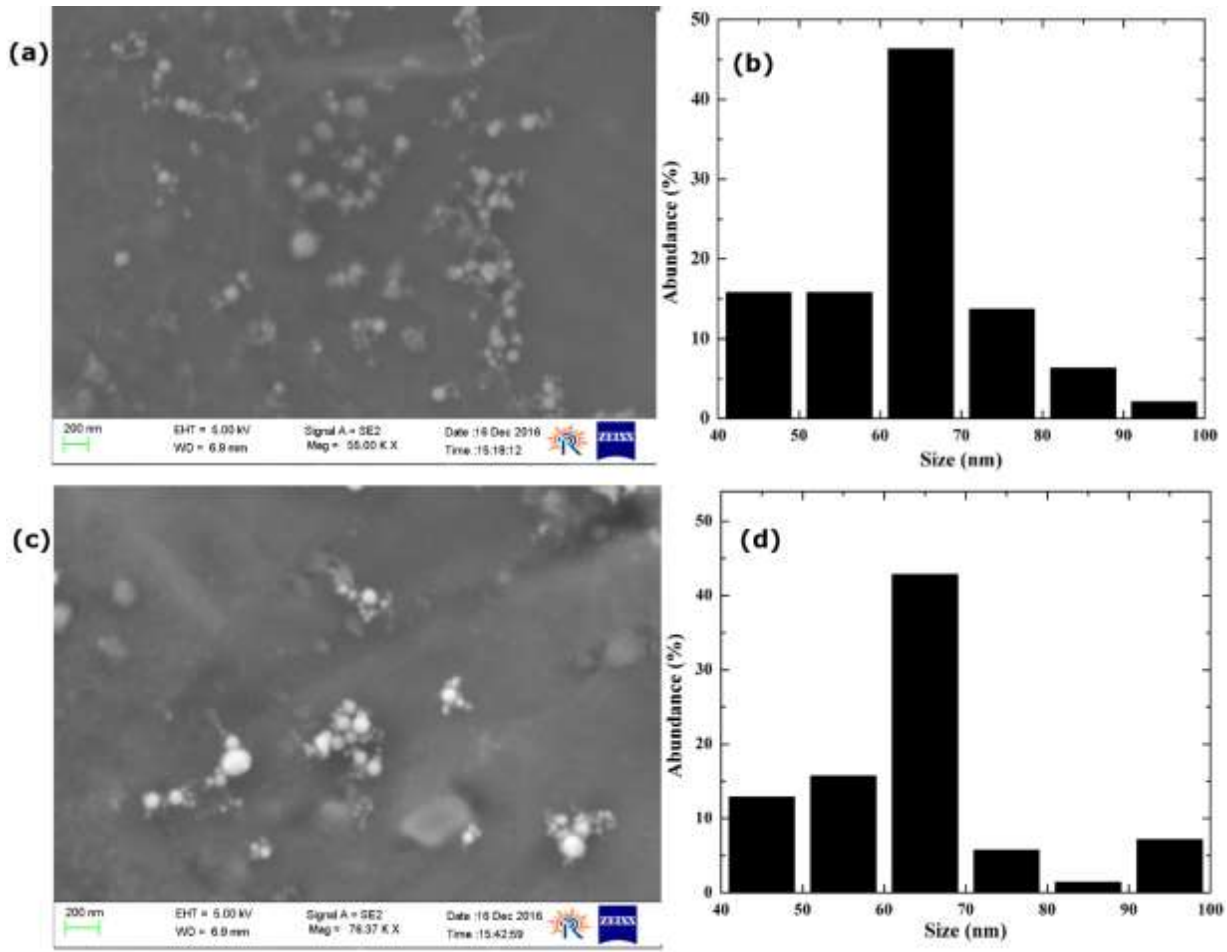

**Figure 8:** SEM micrographs of the nanoparticles produced as a result of ablation of a flat titanium target (a) and titanium target fitted with confining physical boundaries (c). Corresponding size-distributions of the resultant nanoparticles have been presented in (b) and (d). (Ambient liquid medium: IPA)

| Table 4: Comparison of the mean size of the nanoparticles produced with flat target to those obtained with the target fitted with physical boundaries (Ambient liquid medium: IPA) | | | |
|---|---|---|---|
| Material | Mean size of nanoparticles | | % Increment |
| | Flat target | Target with boundary | |
| Copper | 34.5±18 mm | 76.3±17 mm | 121 |
| Titanium | 63.9±12 mm | 70.6±22 mm | 10.5 |

Results of various experiments presented in the above two sections reveal that the changes in the size of the nanoparticles (as one goes from the flat target to the ones fitted with physical boundaries) is significantly more appreciable in the case of copper target material as compared to those observed with titanium target. On the other hand, the particle size distribution is found to be sharper in the case of titanium-based target. This observation may be attributed to the large difference in the thermal conductivities of copper (3.98 $Wcm^{-1}K^{-1}$) and titanium (0.22 $Wcm^{-1}K^{-1}$) target materials. Owing to the relatively lower thermal conductivity of titanium, the amount of thermal energy produced due to laser ablation remains localized in a smaller spatial domain and hence the mass ablated per shot is expected to be more as compared to that in copper and hence an increment in the resulting particle size is to be expected in the case of titanium-based target material. It is pertinent to mention here that one may explain these trends (resultant particle size distributions) on the basis of parameters, like surface energy, electron-phonon coupling strength etc. as well. However, the primary findings of the present experimental work have clearly identified the plausible roles of the presence of physical boundaries on either sides of the ablation site in deciding the size and the size distribution of the nanoparticles produced due to the laser ablation of the target material(s) placed in two different liquid media.

## 5. Conclusions

The plausible control on the size of the nanoparticles produced as a result of ablation of metallic target material using confining physical boundaries on either side of the ablation site was experimentally demonstrated. Experiments were performed with copper and titanium target materials with two different ambient liquid medium of water and isopropyl alcohol (IPA). The study clearly highlighted the potential of employing physical boundaries in manipulating the sizes and size distributions of the produced nanoparticles. The experiments revealed that for any given liquid medium, irrespective of the target material, the mean size of the nanoparticles obtained due to the ablation of the target plate fitted with confining boundaries is consistently higher than that achieved with the flat target materials. Furthermore, a comparison of the mean particles sizes obtained with the two materials revealed that in any liquid medium, the difference in the size of copper nanoparticles produced from flat and walled targets was much higher and sharper as compared to that observed in the case of titanium nanoparticles produced under similar experimental conditions. The observed increase in the size of the nanoparticles in the presence of the confining boundaries was primarily attributed to the prolonged thermalisation of the plasma plume (due to the reflection of the shockwaves). Select experiments with gold as the target material were also performed in water as the liquid medium to ensure that the observed differences in the sizes of the nanoparticles (with and without the confining boundaries) were primarily due to the prolonged thermalisation of the plasma in the presence of physical barriers and not due to the possible formation of the oxide layer. These gold-based experiments supported the proposition made in the present work that physical barriers can be one of the potential methods to control the size and/or the size distribution of the metallic nanoparticles formed due to the phenomenon of laser ablation in liquids. Taking into account the inherent advantages of the proposed methodology over the other conventionally employed techniques (e.g. wet chemical methods), the usage of the physical boundaries on either side of the ablation site presents a novel, yet one of the most simplistic, ways of controlling the size and size distributions of the produced nanoparticles.

**Acknowledgements:** Authors are grateful to Dr. Mukesh Ranjan for providing us with access to the SEM facility.

# Appendix

The algorithm used to determine the size of the nanoparticles from the available SEM images has been shown in the flow-chart below.

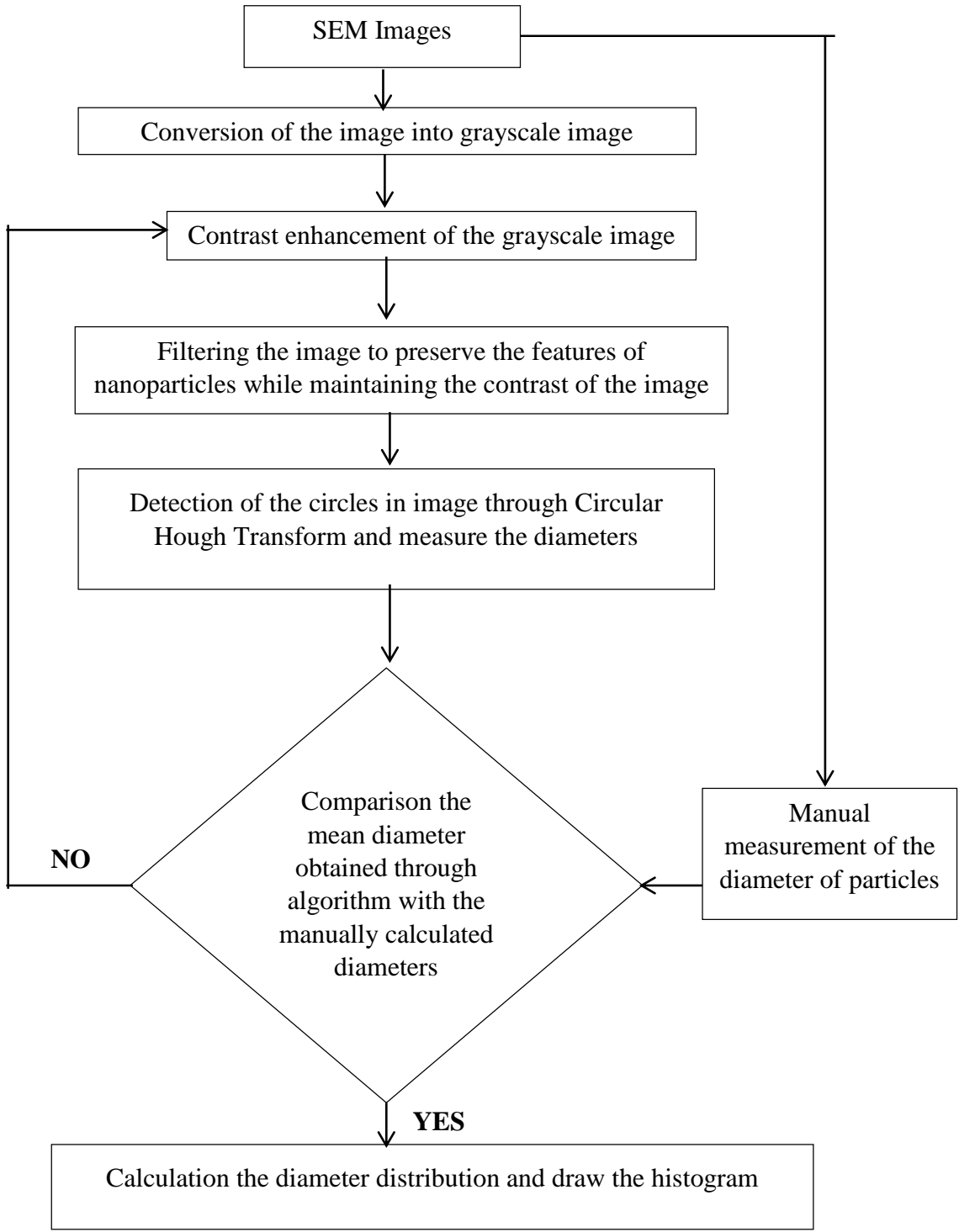

**Figure A-1: Flow-chart of the algorithm employed to extract the sizes of the nanoparticles from the SEM images**